%% file: ttres.tex
\documentclass[twocolumn,epjc3]{svjour3}  
\pdfoutput=1
\smartqed  

\RequirePackage{graphicx}
\usepackage{pstricks}
\usepackage{subfigure}
\usepackage{rotating}
\usepackage{mathrsfs}
\usepackage{hyperref}
\usepackage{dcolumn}
\usepackage{subfigure}
\usepackage{cite}

\newcommand{\lumitot}{\mbox{2.05~fb$^{-1}$}}
\newcommand{\limfive}{\mbox{9.3}}
\newcommand{\limthirteen}{0.95}
\newcommand{\limzpold}{\mbox{860}}
\newcommand{\limzp}{\mbox{880}}
\newcommand{\limkkg}{\mbox{1130}}
\newcommand{\limkkgDIL}{\mbox{1080}}
\newcommand{\limkkgDILexp}{\mbox{1070}}
\newcommand{\pt}{\ensuremath{p_{\mathrm{T}}}} 
\newcommand{\pT}{\ensuremath{p_{\mathrm{T}}}} 
\newcommand{\HT}{\ensuremath{H_{\mathrm{T}}}} 
\newcommand{\met}{\ensuremath{E_{\mathrm{T}}^{\mathrm{miss}}}} 
\newcommand{\HTM}{\mbox{\ensuremath{H_{\mathrm{T}}}+\ensuremath{E_{\mathrm{T}}^{\mathrm{miss}}}}} 
\newcommand{\AKT}{anti-$k_{t}$}
\def\tev{\ifmmode {\mathrm{\ Te\kern -0.1em V}}\else
                   \textrm{Te\kern -0.1em V}\fi}%
\def\gev{\ifmmode {\mathrm{\ Ge\kern -0.1em V}}\else
                   \textrm{Ge\kern -0.1em V}\fi}%
\usepackage{preprintcover}  
\PreprintCoverPaperTitle{A search for $t\bar{t}$ resonances with the ATLAS detector
in 2.05 fb$^{-1}$ of proton-proton collisions at $\sqrt{s} = 7$ \tev}  
\PreprintIdNumber{CERN-PH-EP-2012-122}  
\PreprintCoverAbstract{A search for top quark pair resonances in 
final states containing at least one electron or muon
has been performed with the ATLAS experiment at the CERN Large Hadron Collider.
The search uses a data sample corresponding to an integrated luminosity
of \lumitot{}, which was recorded in 2011 at a proton-proton centre-of-mass energy of
7 \tev. No evidence for a resonance is found and limits are set on the production cross-section times
branching ratio to $t\bar{t}$ for narrow and wide resonances. 
For narrow $Z'$ bosons, the observed 95\% Bayesian credibility level limits range from 
\limfive~pb to \limthirteen~pb for masses in the range of $m_{Z'}=$~500 \gev\ to $m_{Z'}=$~1300 \gev. 
The corresponding excluded mass region for a leptophobic topcolour $Z'$ boson 
(Kaluza-Klein gluon excitation in the Randall-Sundrum model) is $m_{Z'} <$~\limzp\ \gev\
($m_{g_{\rm KK}} <$~ \limkkg\ \gev).}
\PreprintJournalName{European Physical Journal C}  

\journalname{Eur. Phys. J. C}
\begin{document}

\title{A search for $t\bar{t}$ resonances with the ATLAS detector
in 2.05 fb$^{-1}$ of proton-proton collisions at $\sqrt{s} = 7$ \tev}

\author{The ATLAS Collaboration}
\institute{CERN, 1211 Geneva 23, Switzerland E-mail: atlas.publications@cern.ch}

\date{July 7, 2012}

\maketitle

\begin{abstract}
A search for top quark pair resonances in 
final states containing at least one electron or muon
has been performed with the ATLAS experiment at the CERN Large Hadron Collider.
The search uses a data sample corresponding to an integrated luminosity
of \lumitot{}, which was recorded in 2011 at a proton-proton centre-of-mass energy of
7 \tev. No evidence for a resonance is found and limits are set on the production cross-section times
branching ratio to $t\bar{t}$ for narrow and wide resonances. 
For narrow $Z'$ bosons, the observed 95\% Bayesian credibility level limits range from 
\limfive~pb to \limthirteen~pb for masses in the range of $m_{Z'}=$~500 \gev\ to $m_{Z'}=$~1300 \gev. 
The corresponding excluded mass region for a leptophobic topcolour $Z'$ boson 
(Kaluza-Klein gluon excitation in the Randall-Sundrum model) is $m_{Z'} <$~\limzp\ \gev\
($m_{g_{\rm KK}} <$~ \limkkg\ \gev).
\end{abstract}

\section{Introduction}
\label{sec:intro}
The Standard Model of particle physics (SM) is believed to be an effective theory
valid up to energies in the \tev\ range. 
Since particle masses are central to the breaking of the electroweak symmetry, 
final states that involve the heaviest of the
particles presumed to be elementary, the top quark, offer particular promise
in searches for new physics.  This Article describes
searches for new heavy particles decaying to top quark pairs ($t\bar{t}$) using the ATLAS
detector~\cite{Aad:2008zzm} at the CERN Large Hadron Collider (LHC).  
Multiple final state topologies containing at least one lepton (electron or muon) are
considered, in which the lepton is expected to originate from the decay 
of one of the $W$ bosons produced in the top quark decays.
In events with one lepton -- the lepton plus jets ($\ell+{\rm jets}$) channel -- 
the reconstructed $t\bar{t}$ mass spectrum
is used to search for a signal. In events with 
two leptons -- the dilepton channel --
the effective mass is used.
Both variables are defined in 
Section \ref{sec:mreco}.

The benchmark model used to quantify the experimental sensitivity
to narrow resonances is a topcolour $Z'$ boson~\cite{Harris:1999ya} arising
in models of strong electroweak symmetry breaking through top quark
condensation~\cite{Hill:1994hp}. 
The specific model used is the leptophobic scenario, model IV in
Ref.~\cite{Harris:1999ya} with $f_1=1$ and $f_2 =0$ and 
a width of 1.2\%
of the $Z'$ boson mass.  
The model used for wide resonances is a Kaluza-Klein (KK) gluon $g_{\rm KK}$, which appears in 
Randall-Sundrum (RS) models in which particles
are located in a warped dimension~\cite{Agashe:2006hk,Lillie:2007yh,Djouadi:2007eg,Allanach:2009vz}.
The left-handed ($g_L$) and right-handed ($g_R$) couplings to quarks take the 
conventional RS values~\cite{Lillie:2007yh}: 
$g_L=g_R=-0.2 g_s$ for light quarks including charm, where $g_s = \sqrt{4\pi \alpha_s}$;
$g_L=1.0 g_s, g_R=-0.2 g_s$ for bottom quarks; and 
$g_L=1.0 g_s, g_R=4.0 g_s$ for the top quark.
In this case, the resonance width is 15.3\% of its mass, larger
than the detector resolution.  

Previous searches for $t\bar{t}$ resonances were most recently carried out 
by the CDF~\cite{:2007dz,:2007dia,Aaltonen:2009tx,Aaltonen:2011ts,Aaltonen:2011vi} 
and D0~\cite{Abazov:2008ny,Abazov:2011gv} 
collaborations at Run II of the Fermilab Tevatron Collider, and by the 
CMS collaboration~\cite{cmsallhad} at the LHC.
No evidence for new particles was uncovered and 95\% confidence level limits were set on the mass of 
a leptophobic topcolour $Z'$ boson~\cite{Hill:1993hs} at $m_{Z'} > 900$
\gev ~\cite{Aaltonen:2011ts}  as well as
on the coupling strength of a heavy colour-octet vector particle.

\section{The ATLAS detector}
\label{sec:atlas}

The ATLAS detector~\cite{Aad:2008zzm} is designed to measure the properties of particles
produced in proton-proton ($pp$) interactions with excellent precision.  Its cylindrical 
geometry, with axis aligned with the proton beams, is augmented by two endcap sections.
This results in almost complete 4$\pi$ solid
angle coverage.  The Inner Detector (ID) covers pseudorapidities\footnote{
ATLAS uses a right-handed coordinate system with its origin at the nominal interaction 
point (IP) in the centre of the detector and the $z$-axis along the beam pipe. The $x$-axis 
points from the IP to the centre of the LHC ring, and the $y$ axis points upward. 
Cylindrical coordinates $(r,\phi)$ are used in the transverse plane, $\phi$ being the 
azimuthal angle around the beam pipe. The pseudorapidity is defined in terms of the polar 
angle $\theta$ as $\eta=-\ln\tan(\theta/2)$.} of $|\eta | <$ 2.5 and
 consists of layers of silicon pixel and strip 
detectors and a straw-tube transition radiation tracker.
It is embedded in the bore of 
a 2 T superconducting solenoidal magnet to allow precise measurement of charged 
particle momenta.  This system is surrounded by a hermetic calorimeter system consisting 
of finely segmented sampling calorimeters using lead/liquid-argon for the 
detection of electromagnetic (EM)
showers up to $|\eta| <$ 3.2, and copper or tungsten/liquid-argon for 
hadronic showers for 1.5 $< |\eta| <$ 4.9.  In the central 
region ($|\eta| <$ 1.7), an iron/scintillator hadronic calorimeter is used.
Outside the calorimeters, the muon spectrometer incorporates multiple layers of
trigger and tracking chambers within an air-core toroidal magnetic field,
enabling an independent, precise measurement of muon track momenta.

\section{Data sample}
\label{sec:data}

The data were collected with the \mbox{ATLAS} detector at the 
CERN LHC in 2011 using
single-lepton triggers
with transverse momentum thresholds at 20 \gev\ or 22 \gev\ for electrons and 18 \gev\
for muons.  These triggers use similar, but looser selection criteria than
the \mbox{offline} reconstruction and reach their
efficiency plateaus at 25~\gev\ (electrons) and 20~\gev\ (muons).

Only data where all subsystems were operational
are used. Applying these requirements to $pp$
collision data recorded with stable beam conditions between March
and August 2011 at $\sqrt{s}=7$ \tev\ 
results in a data sample of 2.05 $\pm$ 0.08 fb$^{-1}$~\cite{Aad:2011dr,LuminosityGroup:1367408}. 

\section{Simulated samples}
\label{sec:sim}

The irreducible SM $t\bar{t}$ background is simulated using 
 {\sc MC@NLO v3.41}~\cite{Frixione:2002ik,Frixione:2003ei}
with {\tt CTEQ6.6}~\cite{Nadolsky:2008zw} parton distribution functions (PDFs),
interfaced
to {\sc Herwig v6.5}~\cite{Corcella:2000bw} for the parton shower and 
hadronization steps and
{\sc Jimmy}~\cite{Butterworth:1996zw} to model effects due to  the underlying event
and multiple parton interactions. The top quark mass is set to 172.5 \gev\ and 
only events in which at least one of
the $W$ bosons decays leptonically are generated. The inclusive cross-section
of 165 pb is taken from approximate 
next-to-next-to-leading-order (NNLO) calculations~\cite{Moch:2008qy}. 
Electroweak single top quark production is simulated using the same programs,
and cross-sections are based on approximate NNLO
 calculations: 65 pb ($t$-channel)~\cite{Kidonakis:2011wy},
 4.6 pb ($s$-channel)~\cite{Kidonakis:2010tc} and 
 15.7 pb ($Wt$ process)~\cite{Kidonakis:2010ux}.
Samples produced with different parameter settings or other Monte Carlo (MC) 
event generators are used to evaluate the systematic uncertainties due to the top quark mass,
modelling of the shape of the $t\bar{t}$ mass distribution ({\sc Powheg}~\cite{Frixione:2007vw}), 
the parton shower model ({\sc Powheg}+ {\sc Herwig} compared to 
{\sc Powheg}+{\sc Pythia}~\cite{Sjostrand:2006za}), and
initial- and final-state radiation effects (using {\sc AcerMC}~\cite{Kersevan:2004yg}).  
These last uncertainties are considered both separately
and in a correlated way.

Production of a $W$ or $Z$ boson plus jets with leptonic vector boson decays
is simulated with {\sc Alpgen v2.13}~\cite{Mangano:2002ea} and 
{\tt CTEQ6L1}~\cite{Pumplin:2002vw}
PDFs in exclusive
bins of parton multiplicity for multiplicities lower than five, and inclusively
above that. For the $Z$ boson plus jets sample, $Z$-photon interference is included
and events are required to have a dilepton invariant mass in the range
$10 < m_{\ell\ell} < 2000$~\gev. The events are processed by {\sc Herwig}
and {\sc Jimmy}, and matrix-element--parton-shower matching is performed with 
the MLM~\cite{Hoche:2006ph} \\ method.
The inclusive samples are initially normalized to the NNLO 
cross-sections~\cite{Hamberg:1990np,Gavin:2012kw}, and in addition 
later corrected using data as described in Section~\ref{sec:wjets} and Section~\ref{sec:zjets}.

Diboson samples for the $\ell+{\rm jets}$ channel are produced using~{\sc Herwig v6.5} 
with {\sc Jimmy} and \\
{\tt MRST2007LO$^*$}~\cite{Sherstnev:2007nd} PDFs with {\sc Jimmy}. A
filter requires the presence of one lepton with $\pt > $  10~\gev\ and
pseudorapidity $|\eta| <$ 2.8.
The cross-sections used for these filtered
samples are 11.8~pb  for $WW$ production, 3.4~pb  for $WZ$
production, and 0.98~pb  for $ZZ$ production.  
These values are multiplied with ``K-factors'' of 1.52, 1.58 and 1.20,
corresponding to the ratio of the next-to-leading-order (NLO) and leading-order (LO) calculations, and
obtained using the {\sc MCFM}~\cite{MCFM,Campbell:2011bn} generator.
Additional diboson samples
for the dilepton channel are 
simulated using {\sc Alpgen v2.13} with {\tt CTEQ6L1} PDFs and interfaced with {\sc Herwig} 
and {\sc Jimmy}. 

Signal samples for $Z'$ bosons decaying to $t\bar{t}$ are generated using 
{\sc Pythia v6.421} with {\tt CTEQ6L1} PDFs allowing all top quark decay modes.  
Cross-sections for the $Z'$ boson samples are evaluated 
with an updated calculation~\cite{Harris:2011ez}
to which a K-factor of 1.3 is applied~\cite{Gao:2010bb}.
Samples of KK gluons are generated with {\sc Madgraph} \newline {\sc v4.4.51}~\cite{Alwall:2007st}, 
and showered with {\sc Pythia} without taking into account interference with 
SM $t\bar{t}$ production, and
the cross-sections are recalculated using {\sc Pythia v8.1}~\cite{Sjostrand:2007gs}.  
In both cases, {\tt CTEQ6L1} PDFs are used.  
The resulting cross-sections are given in Table~\ref{tab:signals}.
\begin{table}[tbh]
\begin{center}
\caption{Cross-sections times branching ratios for the resonant signal processes obtained using the 
generator and PDF combinations described in the text.  The KK gluon ($Z'$) cross-sections are given 
at LO (LO $\times$ 1.3).}
\vspace{2mm}
\begin{tabular}{lrr}
\hline
Signal mass [\gev] & \multicolumn{2}{c}{$\sigma \times$ BR($Z'/g_{\rm KK} \to t\bar{t}$) [pb]}\\ \hline
 & Topcolour $Z'$ & $g_{\rm KK}$ \\ \hline 
500 \gev & 19.6 & 81.2 \\
600 \gev & 10.3 & 39.4 \\
700 \gev & 5.6 & 20.8 \\
800 \gev & 3.2 & 11.6 \\
900 \gev & 1.9 & 6.8 \\
1000 \gev & 1.2 & 4.1 \\
1200 \gev & 0.46 & 1.7 \\
1400 \gev & 0.19 & 0.73 \\
1600 \gev & 0.086 & 0.35 \\
1800 \gev & 0.039 & 0.18 \\
2000 \gev & 0.018 & 0.095 \\
\hline
\end{tabular}
\label{tab:signals}
\end{center}
\end{table}

After event generation, all samples are processed by
a {\sc GEANT4}-based~\cite{Agostinelli:2002hh} simulation of the 
ATLAS detector~\cite{:2010wqa} and reconstructed using the same software as
used for data.
All simulated samples include the effects
due to multiple $pp$ interactions per bunch-crossing, and events are reweighted so
that the data and simulated sample instantaneous luminosity profiles match.

\section{Object reconstruction}
\label{sec:reco}

Electron candidates must have an EM shower shape consistent with 
expectations based on simulation, test-beam and $Z \to ee$ events 
in data, and must have a matching track in the ID~\cite{Aad:2011mk}.
They are required to have transverse momentum $\pT >25$~\gev\ and
$|\eta_{\mathrm{cluster}}|<2.47$, 
where $\eta_{\mathrm{cluster}}$ is the pseudorapidity of the calorimeter
cluster associated to the candidate.
Candidates in the calorimeter transition
region at $1.37<|\eta_{\mathrm{cluster}}|<1.52$ are excluded.

Muon candidates  are reconstructed from track segments in the
various layers of the muon chambers, and matched with tracks
found in the ID. The final candidates are refitted using
the complete track information from both detector systems, and
required to satisfy $\pT>25$~\gev\ and $|\eta|<2.5$.
Additionally, muons are required to be separated by $\Delta R > 0.4$
from any jet with $\pT>20$~\gev.

The leptons in each event
are required to be isolated~\cite{Aad:2012qf}
to reduce the background due to non-prompt leptons, e.g. from decays of hadrons
(including heavy flavour) produced in jets.
For electrons, the calorimeter isolation transverse energy in a cone  
in $\eta$-$\phi$ space of radius $\Delta R=0.2$ 
around the electron 
position\footnote{The radius $\Delta R$
between the object axis and the edge of the object cone is defined as
$\Delta R = \sqrt{({\Delta \phi})^2+({\Delta \eta})^2}$.}
is required to be less than 3.5~\gev.
The core of the electron energy deposition is excluded and the sum is corrected for transverse shower
leakage and pile-up from additional $pp$ collisions.
For muons, the calorimeter isolation transverse energy, corrected for muon energy deposition, 
in a cone of
$\Delta R=0.3$ is required to be less than $4.0$~\gev.  The scalar sum
of track transverse momenta in a cone of $\Delta R=0.3$ around but
excluding the muon track is also
required to be less than $4.0$~\gev.

Jets are reconstructed with the \AKT\ algorithm~\cite{Cacciari:2005hq,Cacciari:2008gp} 
with radius parameter $R=0.4$ from topological clusters~\cite{Aad:2011he} of energy deposits in
the calorimeters, calibrated at the EM energy scale
appropriate for the energy deposited by electrons or photons.  
These jets are then calibrated to the hadronic energy scale, using a
$\pT$- and $\eta$-dependent correction factor~\cite{Aad:2011he} obtained from simulation, 
test-beam and collision data.
The uncertainty on this correction factor is determined
from control samples in data.  
Jets must have $\pT >20$~\gev\ and $|\eta| < 4.5$.
If the closest object to an electron candidate
is a jet with a separation $\Delta R<0.2$ the jet is removed in order
to avoid double-counting of electrons as jets.
While the topological clusters are taken to be massless, 
jets are composed of many of these, and their
spatial distribution within the jet cone leads to an 
invariant mass~\cite{Collaboration:2012ef}.

Jets originating from $b$-quarks are selected by exploiting the long
lifetimes of bottom hadrons (about 1.5 ps) leading to typical flight
paths before decay of a few millimeters, which are observable in the detector. A
multivariate $b$-tagging algorithm~\cite{ATLAS-CONF-2011-102} 
is used in this analysis at an operating point yielding,
in simulated $t\bar{t}$ events, an average 60\%
$b$-tagging efficiency and a light quark jet rejection factor of 345.

The missing transverse momentum (\met) is constructed~\cite{Aad:2012re} from the vector sum of
all calorimeter cells contained in topological clusters. 
Calorimeter cells are associated with a parent physics object in a
chosen order: electrons, jets and muons, such that a cell is uniquely
associated to a single physics object. Cells belonging
to electrons are calibrated at the electron energy scale, but omitting
the out-of-cluster correction to avoid double cell-energy counting,
while cells belonging to jets are taken at the corrected energy scale
used for jets.  Finally, the \pt\ of muons passing
selection requirements is included, and the contributions from any
calorimeter cells associated to the muons are subtracted. The
remaining energy clusters not associated to electrons or jets are
included at the EM scale.

For all reconstructed objects in simulation, scaling factors are applied to 
compensate for the difference in reconstruction efficiencies between data
and simulation.  The uncertainties on these scaling factors are used to 
determine the corresponding systematic uncertainties.

\section{Event selection}
\label{sec:evtsel}

After the event has been accepted by the trigger, it is required to have at least one
offline-reconstructed primary vertex with at least five tracks with \pt $> 0.4$ \gev, and it
is discarded if any jet with $\pT>20$~\gev\ is
identified as out-of-time activity or calorimeter noise~\cite{Aad:2011he}.

\subsection{$\ell+{\rm jets}$ channel}

The event must contain exactly one isolated lepton, and 
events where an electron shares an inner detector
track with a non-isolated muon, or with a second lepton with $\pt >$
15~\gev, are rejected. 
The total
$t\bar{t}$ event fraction is enhanced by applying the following event-level cuts.
In the electron channel, \met{} must be larger than 35~\gev\ and 
 $m_{\rm T} > 25$~\gev, 
 where $m_{\rm T}$ is the 
 lepton-\met\ transverse mass\footnote{The transverse mass is defined by 
the formula $m_{\rm T} = \sqrt{2\pt^\ell \met (1-\cos\Delta\phi)}$, where $\pt^\ell$
is the lepton \pt\ and $\Delta \phi$ is the azimuthal angle between the lepton and \met.};
in the muon channel, \met{} $> 20$~\gev\ and \met{}+$m_{\rm T} > 60$~\gev\ are required.
If one of the jets has mass $m_j > 60$~\gev, the event must contain at least three jets with
 $\pt>25$~\gev\ and $|\eta|<2.5$;  if not, at least four 
 jets satisfying the same \pt\ and $\eta$ criteria must be present.
The leading jet must have $\pt>60$~\gev, and at least one of the jets must be tagged as a $b$-jet.
The requirement on the number of jets is relaxed when one jet has $m_j > 60$~\gev\ since
for top quarks with significant boost the decay products are collimated, and multiple
quarks from top quark or $W$ boson decay can be reconstructed as a single, massive jet.
This subsample represents approximately 0.3\% of the selected event sample.
The total signal acceptance times branching ratio to $t\bar{t}$ is 
is 7.4\% for a topcolour $Z'$ boson of mass $m_{Z'}=800$~\gev\ 
and 7.3\% for a KK-gluon of mass $m_{g_{\rm KK}}=1300$~\gev.

\subsection{Dilepton channel}
The event selection follows that used in a recent ATLAS $t\bar{t}$ production 
cross-section measurement~\cite{ATLAS:2012aa}.  Candidate events 
are required to have two isolated leptons of opposite charge and
two or more jets with $\pt>$~25~\gev. In order to suppress the $Z$ plus jets background, 
$ee$ and $\mu\mu$ events are required 
to have an invariant dilepton mass outside the $Z$ boson mass window, defined as 
$|m_{Z}-m_{\ell\ell}| < 10$~\gev, and $\met > 40$~\gev. An additional cut  
$m_{\ell\ell} > 10$~\gev\ is applied to the data in order to conform with the lower 
$m_{\ell\ell}$ cut-off in the $Z$ plus jets simulation and to reduce backgrounds from meson 
resonances. In the $e\mu$ channel the non-$t\bar{t}$ background is suppressed by 
requiring the scalar sum of the transverse momenta of the identified leptons and jets to be 
larger than $130$~\gev. 
The total signal acceptance times branching ratio to $t\bar{t}$ is 
1.3\% for a topcolour $Z'$ boson of mass $m_{Z'}=800$~\gev\
and 1.5\% for a KK-gluon of mass $m_{g_{\rm KK}}=1100$~\gev.

\section{Data-driven background modelling}
\label{sec:bckgrd}

For the dominant background sources, $t\bar{t}$ and single top production, $W$ plus jets
in the $\ell+{\rm jets}$ channel and $Z$ plus jets in the dilepton channel, 
the simulated samples are corrected based
on measurements in data.  The multijet background is determined directly from data.
All other backgrounds are taken without modification from simulation.

\subsection{SM $t\bar{t}$ and single top modelling}
\label{sec:ttbar}

As discussed in Section~\ref{sec:sim}, 
the SM $t\bar{t}$ and single top backgrounds are simulated using the {\sc MC@NLO} 
generator with {\tt CTEQ6.6} PDFs.
To investigate the impact of the choice of PDFs on modelling of this dominant
background, the events are re-weighted to {\tt MSTW2008nlo}~\cite{Martin:2009iq} PDFs
and the data are compared to the background expectation for angular variables:
jet and lepton rapidities, and azimuthal angles between these objects and \met.
Since the use of {\tt MSTW2008nlo} leads to better agreement in these angular variables, 
samples re-weighted to these PDFs are used in the analysis.  Distributions obtained
with {\tt CTEQ6.6} PDFs are used to estimate the systematic uncertainty associated 
with this shape modelling.

\subsection{$W$ plus jets corrections}
\label{sec:wjets}

For the $\ell+{\rm jets}$ channel, the $W$ plus jets background is determined using 
the {\sc Alpgen} samples described in Section~\ref{sec:sim}, with data-driven corrections.

The flavour composition is determined from data based on the tagged fraction of $W$ plus one- and two-jet 
events~\cite{Aad:2011kp}, and the known $b$-tagging efficiencies, measured using 
various techniques involving jets containing muons~\cite{ATLAS-CONF-2011-089}.
The MC predictions for different flavour contributions are scaled accordingly, 
adjusting the ``light parton'' scale factor to keep the untagged $W$ plus two jets 
normalization unchanged. 
The $Wb\bar{b}$ and $Wc\bar{c}$ components
are scaled by a factor 1.63, the $Wc$ component by a factor 1.11, 
and the ``light parton'' component by a factor 0.83. 
The flavour composition uncertainty of the $W$ plus jets background is estimated by varying these 
scaling factors by their uncertainties (13\% for $Wb\bar{b}$ and $Wc\bar{c}$, 
9\% for $Wc$).  

Normalization 
factors are derived based on the charge asymmetry in $W$ boson production
at the LHC~\cite{Aad:2012ug}:
\[ (N_{W^+} + N_{W^-})^{\rm exp} = \left( \frac{r_{\rm MC}+1}{r_{\rm MC}-1} \right) 
(N_{W^+} - N_{W^-})^{\rm data} \]
where $N_{W^+}$ and $N_{W^-}$ are the number of events with $W^+$ and $W^-$
bosons, $r_{\rm MC} = N_{W^+}/N_{W^-}$, and the superscripts ``$\rm exp$'' and 
``$\rm data$'' denote expected
and data events, respectively.  The difference 
$(N_{W^+} - N_{W^-})^{\rm data}$ and ratio $r_{\rm MC}$ 
are extracted from data and simulation,
respectively, as a function of the number of $b$-tags and the number of 
reconstructed jets passing the selection cuts.
The background contamination in the $W$ boson samples extracted from data is verified 
to be charge-symmetric within uncertainties, and cancels in the difference.
In the tagged four-jet bin, an overall normalization factor for the simulated samples
of $0.91$ ($0.81$) is required in the electron (muon) channel to match the data-driven
prediction.  
The overall normalization
uncertainty on the $W$ plus jets background is set at 48\%, based 
on an uncorrelated, 24\%--per-jet uncertainty with respect to the inclusive 
$W$ boson production cross-section~\cite{Alwall:2007fs}.

\subsection{$Z$ plus jets corrections}
\label{sec:zjets}
Even though the event selection in the dilepton channel includes cuts to reject $Z$ plus jets events, 
a small fraction of events 
in the \met\ tails and dilepton invariant mass sidebands remain.
To estimate this background contribution, the number of Drell-Yan events is
measured in a data control sample orthogonal to the signal sample~\cite{ATLAS:2012aa}.
The control sample consists of events with at least two jets, a dilepton invariant mass inside 
the $Z$ boson mass window, and $\met >40$~\gev.

A small contamination in the control sample from non-$Z$-boson processes is subtracted from data using
simulation. A scale factor is then derived based on {\sc Alpgen} $Z$ plus jets 
samples to extrapolate the
data-to-MC differences measured in the control region (CR) into the signal region (SR):
\[N_{Z{\rm +jets}}^{\rm SR} = \\
\frac{(\mathrm{Data}^{\rm CR} - \mathrm{MC}_{\mathrm{other}}^{\rm CR})}{\mathrm{MC}_{N_{Z{\rm +jets}}}^{\rm CR}}
\mathrm{MC}_{N_{Z{\rm +jets}}}^{\rm SR} \]
where $\mathrm{MC}_{N_{Z{\rm +jets}}}^{\rm SR/CR}$ represents the expected number of 
events in the signal and control regions, respectively. 
$\mathrm{MC}_{\mathrm{other}}^{\rm CR}$ is the number of events from 
non-$Z$ contamination in the control region. $\mathrm{Data}^{\rm CR}$ is the 
observed number of events in the control region.
The $Z$ plus jets background normalization prediction from the simulation
is thus scaled by the ratio of data 
to simulated events in the control region.
In the $\ell+{\rm jets}$ channel the background from 
$Z$ plus jets production is small and evaluated directly from the simulation.

\subsection{Multijet background estimation}
\label{sec:qcd}

Jets, including those containing a leptonically decaying bottom or charmed hadron, 
can fake the isolated lepton signature produced by vector boson decays.  Multijet 
events can thus contain objects that pass the lepton selection
but are not leptons from vector boson decays, and contribute to the selected events.
In the $\ell+{\rm jets}$ channel, the multijet background
expectation and kinematic distributions are determined using the
method
described below. 
It models the multijet
background with a data-driven template, which is normalized
in the multijet-dominated low \met\ region.
Since the multijet background in the 
$b$-tagged samples is dominated by true, non-prompt leptons from heavy flavour quark
decays in both electron and 
muon samples, the template is used for both samples. 

Events for the template are selected from a jet-triggered sample
where exactly one jet with a high electromagnetic fraction (between 0.8 and
0.95) is present. This jet, which in addition must have at least four tracks to
reduce the contribution from photon conversions, is used to model the
lepton candidate. Events in which a good electron candidate is present are rejected,
yielding a sample highly enriched in multijet background with
kinematic characteristics very similar to the multijet events that do pass all
the lepton selection cuts.

To determine the normalization of the multijet background, the data-driven multijet
template and the simulated $t\bar{t}$, single top, $W$ plus jets and $Z$ plus jets
background samples are fitted to the data using the full \met\ spectrum,
i.e.~applying all selections except the \met\ cut. Other contributions are
negligible after all selection cuts. For MC samples,
each bin is allowed to vary according to a Gaussian
distribution centred at the bin height, with 10\% RMS to account for their
own modelling uncertainties.  
The multijet background and signal \met\ spectra are sufficiently different 
so that fitting the multijet contribution to the full distribution 
will not mask a potential signal.  
The multijet template is determined before $b$-tagging to
reduce statistical fluctuations.  The
kinematic distributions in both tagged and untagged samples have been verified 
to agree in shape within the available statistics in data. 

In the dilepton channel, the small multijet background contribution 
is estimated from data using
the Matrix Method \cite{Aad:2011yb}, which accounts for small backgrounds with both 
one ($W$ plus jets background) and two objects (multijet background) mimicking 
leptons from vector boson decays.

\section{Mass reconstruction}
\label{sec:mreco}

\subsection{$\ell+{\rm jets}$ channel}
\label{sec:mrecolj}

To reconstruct the $t\bar{t}$ invariant mass, the neutrino's
longitudinal momentum ($p_z$) is determined by imposing the $W$ boson
mass constraint. If the discriminant of the quadratic equation is negative,
a situation usually due to \met\ resolution effects, the smallest changes to the
\met\ $x$ and $y$ components that lead 
to a null discriminant are applied~\cite{chwalekphd}, leading to an improved
resolution for those two components.
If there are two solutions, the smallest $p_z$ solution is chosen.

Different mass reconstruction algorithms are used for the samples with or without a
jet with $m_j > $ 60~\gev.  In the sample without such a jet,
the dominant source of long, non-Gaussian tails in the mass resolution is the inclusion of a
jet from initial- or final-state radiation in place of one of the jets directly 
related to a top quark decay product.  To reduce this contribution, the
four leading jets with \pt\ $>$ 20~\gev\ 
and $|\eta| <$ 2.5 are considered, and a jet is excluded if its
angular distance to the lepton or closest jet satisfies
$\Delta R > 2.5 - 0.015 \times (m_j/\gev )$.
If more than one jet satisfies this condition, the jet with the largest $\Delta R$
is excluded. 
If a jet was discarded and more than three jets remain, the procedure is iterated.
Then $m_{t\bar{t}}$ is reconstructed from the lepton, \met\ and the leading four jets,
or three jets if only three remain.
The $\Delta R$ cut removes
jets that are well-separated from the rest of the activity in the event.  
Furthermore, by requiring only three jets in the mass reconstruction, the method
allows one of the jets from top quark decay to be outside the detector acceptance, 
or merged with another jet.  

For events with high $t\bar{t}$ mass, the top quark and $W$ boson momenta can be large enough for some
of the decay products to be merged into a single jet, in which case using the four 
highest \pt\ jets often leads to a significant overestimation of $m_{t\bar{t}}$, causing
a substantial contribution to the very high mass tail.  To mitigate 
this, if one of the jets has mass $m_j > $ 60~\gev, it is
combined with the jet closest to it (in $\Delta R$) with $\pt >$ 20~\gev\
to form the hadronic top quark candidate, and the 
other top quark is formed by combining the reconstructed leptonic $W$ boson 
candidate with, among those remaining, the jet with $\pt >$ 20~\gev\ closest to it.  

\begin{figure}[htbp]
\centering
\includegraphics[width=0.45\textwidth]{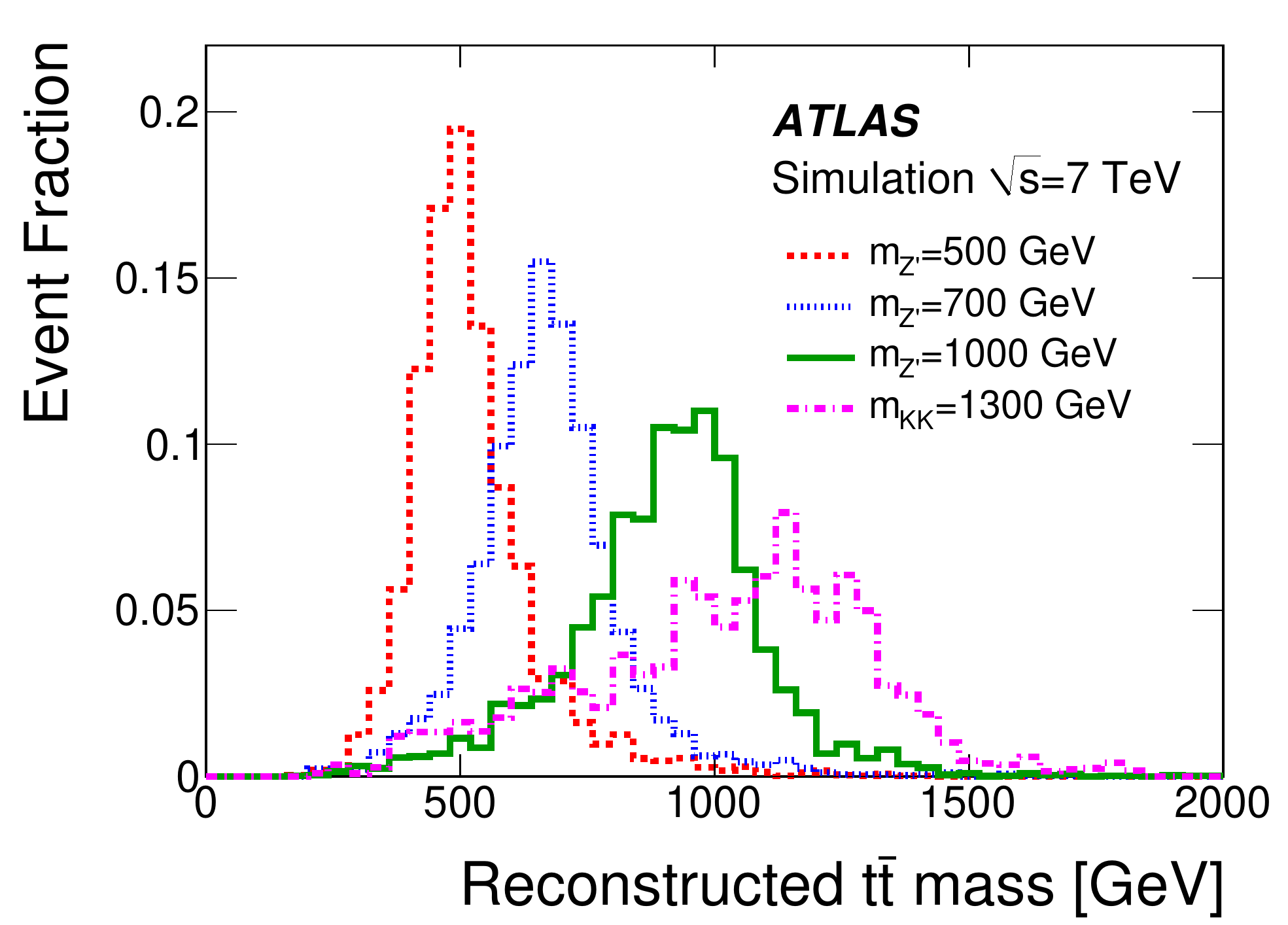}
\caption{Reconstructed $t\bar{t}$ pair invariant mass in simulation
for four resonance masses: $m_{\rm Z'} =$ 500, 700, 1000 and $m_{g_{\rm KK}} =$ 1300~\gev.}
\label{fig:appdrres}
\end{figure}

\begin{figure}[htbp]
\centering
\subfigure[]
{
\includegraphics[width=0.45\textwidth]{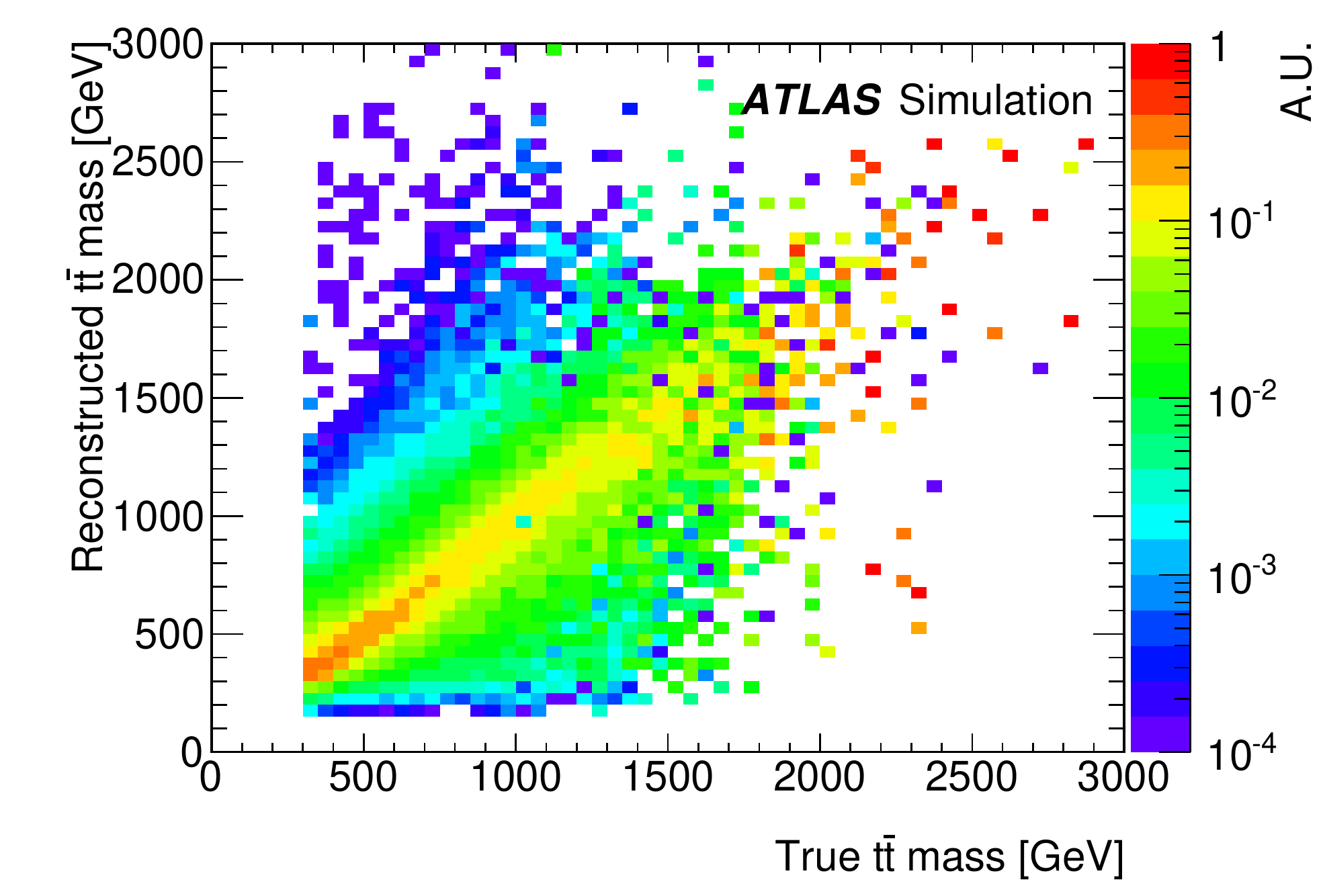}
}
\subfigure[]
{
\includegraphics[width=0.46\textwidth]{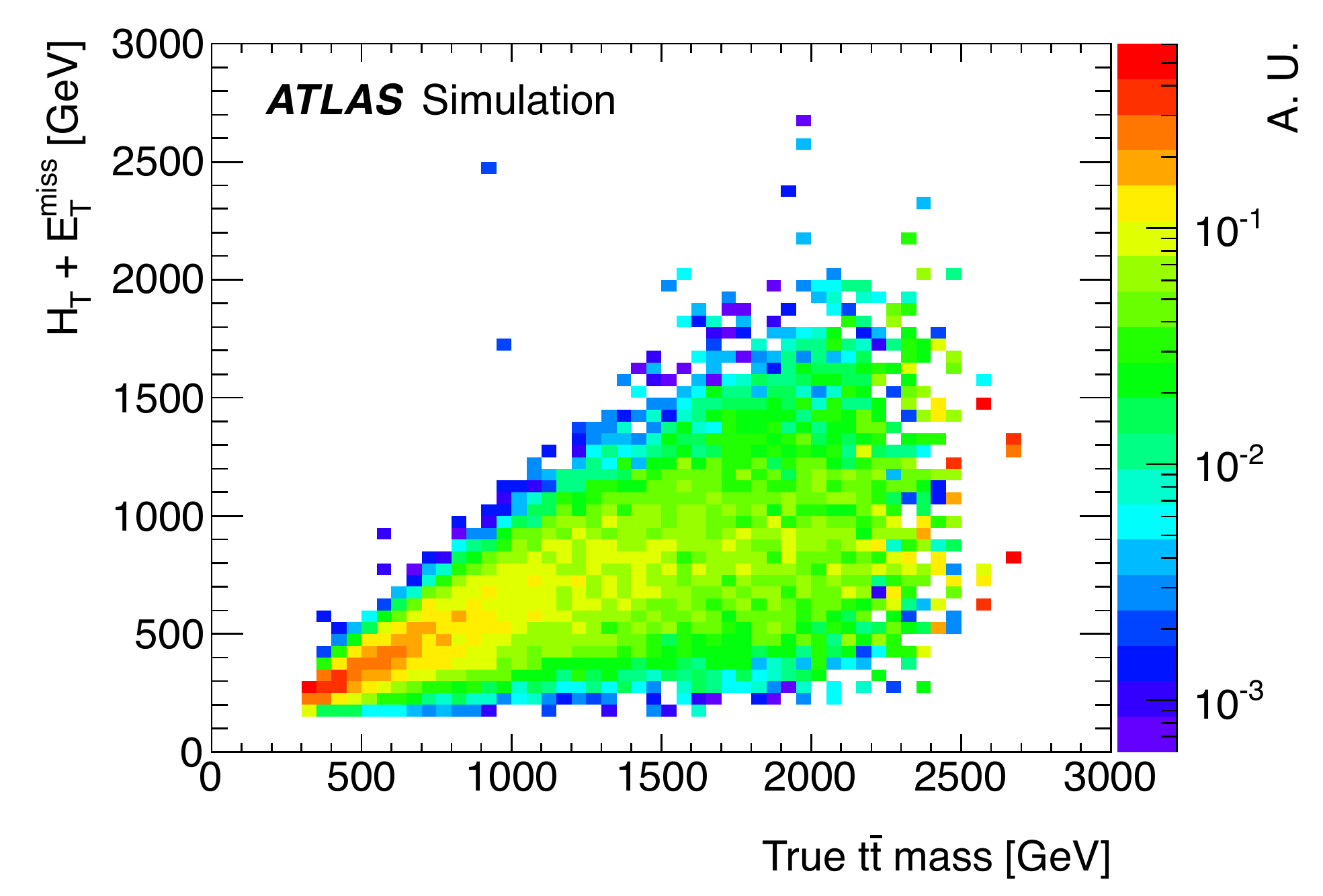}
}
\caption{(a) Reconstructed versus true $t\bar{t}$ pair invariant mass in
the $\ell+{\rm jets}$ channel and (b) effective mass (\HTM) 
versus true $t\bar{t}$ invariant mass
in the dilepton channel. 
The spectrum is normalized to unity for each bin in the true $t\bar{t}$ mass  
to show the 
correlation over a large mass range better.}
\label{fig:drmrvt}
\end{figure}

The mass resolution obtained from simulation is shown in Fig.~\ref{fig:appdrres} using a few 
signal masses, 
and the correlation between true and reconstructed $t\bar{t}$ mass ($m_{t\bar{t}}$) 
is shown in Fig.~\ref{fig:drmrvt}(a).

\subsection{Dilepton channel}
\label{sec:mrecoll}
The dilepton channel is kinematically underconstrained due to 
the presence of two undetected neutrinos. The effective mass is correlated with $m_{t\bar{t}}$
and is defined as \HTM, where \HT\ is the scalar sum of transverse momenta of the leptons
and the two leading jets.
The correlation between true $t\bar{t}$ mass and reconstructed \newline \HTM\ 
is shown in Fig.~\ref{fig:drmrvt}(b).

\section{Systematic uncertainties}
\label{sec:sys}

Since the search for resonances is done using binned $m_{t\bar{t}}$ and \HTM\ distributions,
two categories of systematic uncertainties are considered: uncertainties in the normalization 
of the expected event yield, which do not impact the shapes of the different contributions, 
and uncertainties affecting the shape of the $m_{t\bar{t}}$ or effective mass
distributions, which can also impact the event yields.

Systematic uncertainties that affect only the normalization of the different backgrounds come from 
the uncertainty on the integrated luminosity (3.7\%); 
the lepton trigger and reconstruction efficiencies ($\leq$~1.5\%); and
background normalizations:
$t\bar{t}$ ($^{+7.0}_{-9.6}$\%~\cite{Moch:2008qy}), single top (10\%), diboson (5\%), $W$ or 
$Z$ plus jets in the $\ell+{\rm jets}$ channel (48\%), $Z$ plus jets in the dilepton channel (12\%),
$W$ plus jets and multijet in the dilepton channel (76\%),
multijet in the $\ell+{\rm jets}$ channel (50\%). 
 
The dominant uncertainties that affect both yields and shape
in the $\ell+{\rm jets}$ channel arise from the $b$-tagging 
efficiency~\cite{ATLAS-CONF-2011-089}, with 13\% (17\%) variation in 
the background ($m_{Z'} = $ 800~\gev\ signal) yields, 
jet energy scale including pile-up effects, 15\% (4\%)~\cite{Aad:2011he}, 
and modelling of initial- and final-state radiation, 7\% (6\%).  
The first two have been determined from data by comparing 
results from different methods and/or data samples, while the last has been estimated from 
MC simulations in which the relevant parameters were varied~\cite{Aad:2009wy}.

The largest shape uncertainties in the dilepton channel arise from 
the modelling of initial- and final-state radiation, with 1.0\% (5.1\%) variation in 
the background ($m_{g_{\rm KK}} = $ 1000~\gev\ signal) yields, the jet energy scale 2.5\% (3.0\%) 
and PDFs 3.7\% (0.6\%).

Other uncertainties arising from MC modelling as well as object identification and momentum
measurements have smaller impact.  These include the following: jet energy resolution and
reconstruction efficiency, muon \pt\ resolution, electron energy scale and energy resolution,
\met\ measurement, $m_{t\bar{t}}$ shape (as evaluated by comparison of {\sc Powheg} with 
{\sc MC@NLO}), parton shower and fragmentation ({\sc Pythia} versus {\sc Herwig}), $W$ plus jets
shape (evaluated by varying {\sc Alpgen} generation parameters), $W$ plus jets composition
(from the uncertainty in $Wc$ and $Wc\bar{c}+Wb\bar{b}$ fractions), mis-modelling of 
the multijet background
shape, as well as potential effects due to mis-modelling of pile-up effects.

\section{Comparison of data and background expectation}
\label{sec:datavmc}

\begin{figure}[htbp]
\centering
\includegraphics[width=0.48\textwidth]{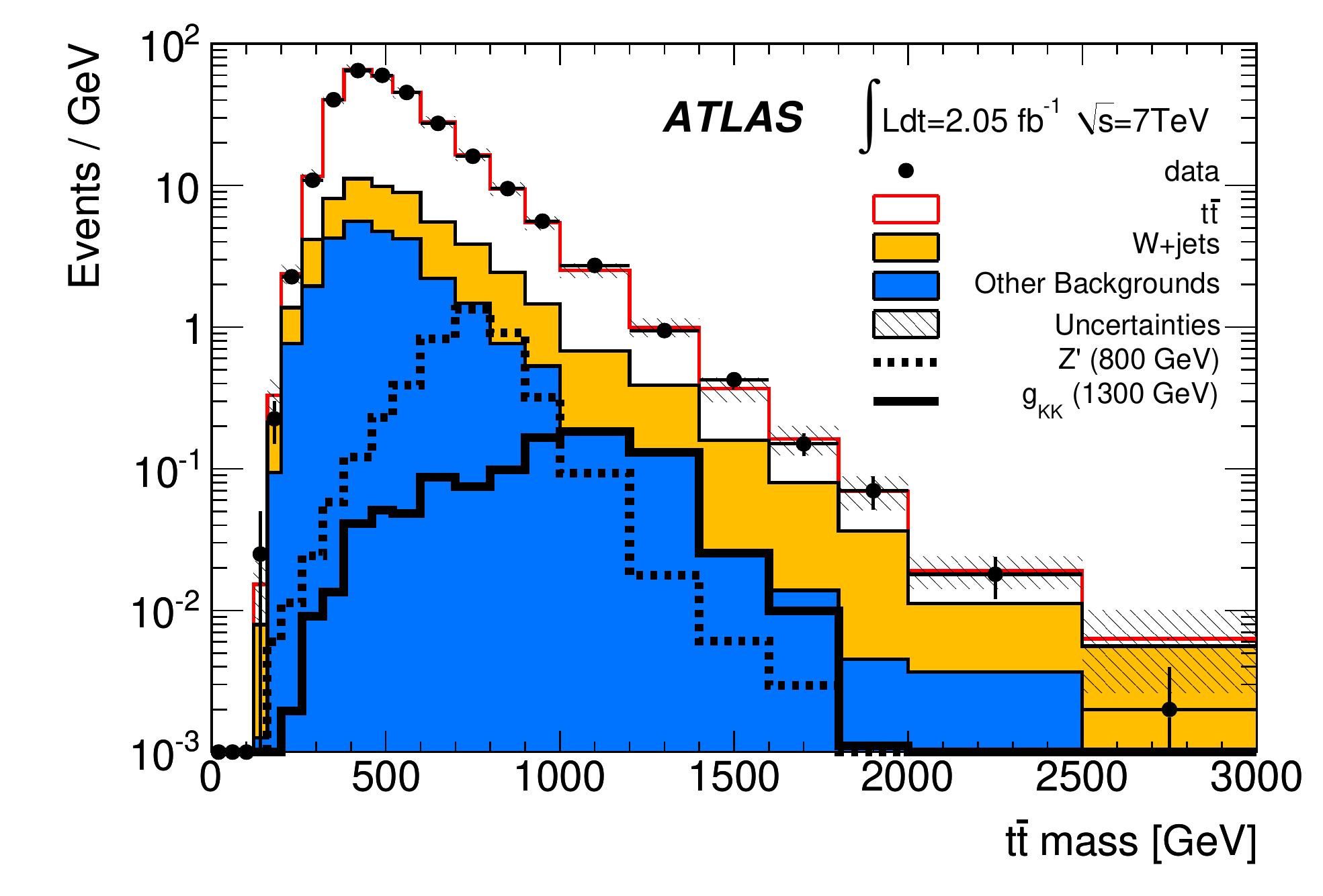}
\caption{Reconstructed $t\bar{t}$ mass in the $\ell+{\rm jets}$ channel
after all cuts, with the expectation from SM background and two signal masses,
a $Z'$ boson with $m_{\rm Z'}$ = 800 \gev\ and a KK gluon with 
$m_{g_{\rm KK}}$ = 1300 \gev.  The electron and muon 
channels have been added together and all events beyond the range of the histogram
have been added to the last bin.  ``Other backgrounds'' includes single top, $Z$ plus jets,
diboson and multijet production.  The hatched area
shows the background normalization uncertainties.}
\label{fig:ctrl-mass}
\end{figure}

\begin{figure}[h]
  \centering
  \includegraphics[width=0.48\textwidth]{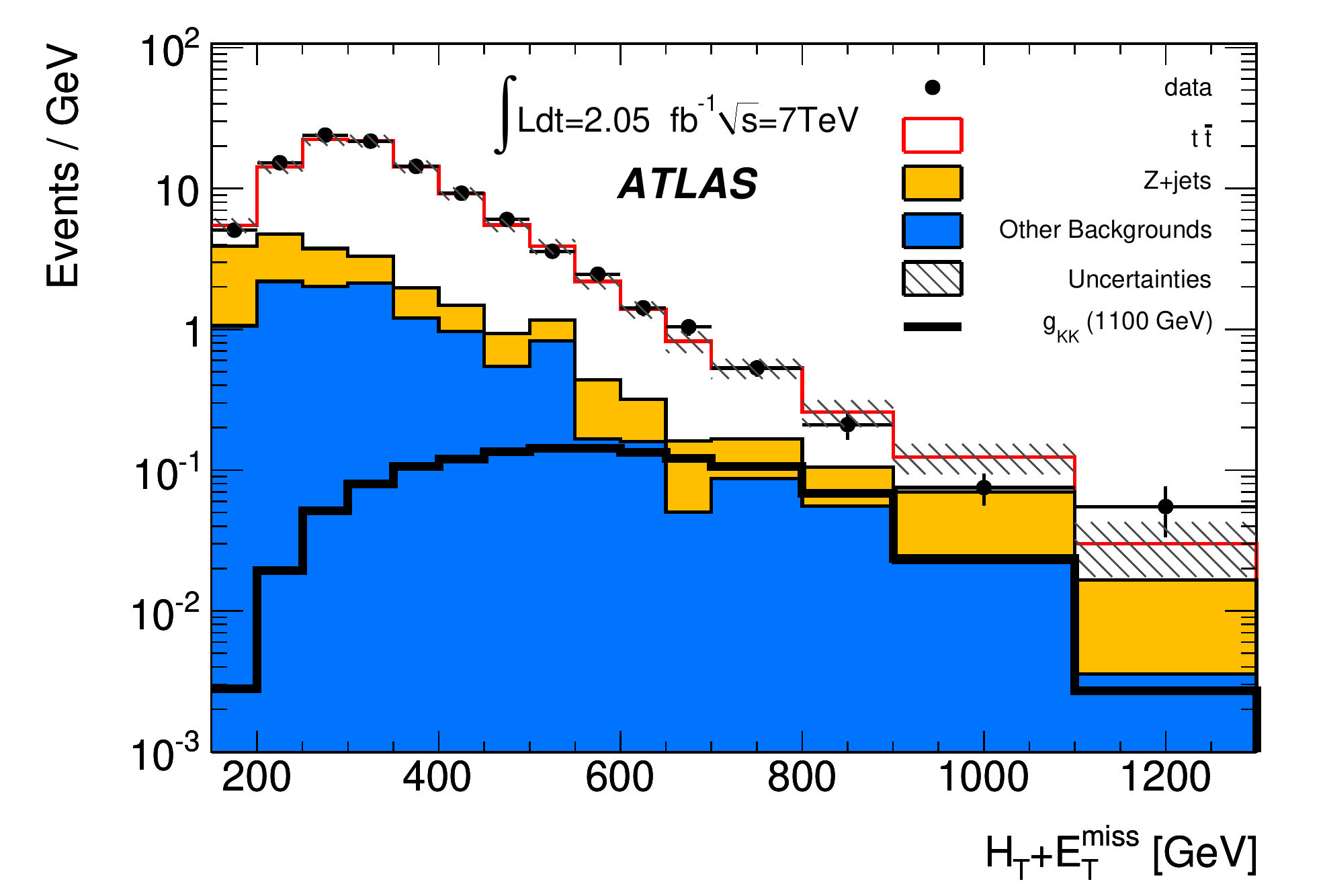}
  \caption{The \HTM~distribution after all selection requirements in the dilepton channel
  with a KK-gluon signal of mass $m_{g_{\rm KK}}=$1100~\gev\ for comparison.
  ``Other backgrounds'' includes single top, diboson, $W$ plus jets,
  and multijet production.
  The hatched area shows the background normalization uncertainties.}
  \label{fig:htscan}
\end{figure}

Tables~\ref{tab:yield} and \ref{tab:dilepton_yield}  compare the predicted and observed 
event yields after applying the event selection cuts described in Section~\ref{sec:evtsel}
for the $\ell+{\rm jets}$ and dilepton channels, respectively. 
\begin{table}[t!]
   \begin{center}
       \caption{
         Number of expected and observed events for the
         $e$ and $\mu+{\rm jets}$ channels after applying all selection cuts
         described in Section~\ref{sec:evtsel}.  The uncertainties 
         given are the normalization uncertainties as described in Section~\ref{sec:sys}.
         Statistical uncertainties on these numbers are small. 
       }
       \label{tab:yield}
       \begin{tabular}{lD{,}{\,\,}{-1}D{,}{\,\,}{-1}} 
       \hline\noalign{\smallskip}
          & \multicolumn{1}{c}{Electron channel} & \multicolumn{1}{c}{Muon channel} \\
       \hline\noalign{\smallskip}
       $t\bar{t}$ & 7830, \pm 750 & 10000, \pm 960 \\
       Single top & 470,\pm 50 & 570,\pm 60 \\
       $W$ plus jets & 1120,\pm 540 & 1450,\pm 700 \\
       $Z$ plus jets & 85,\pm 40 & 90,\pm 45 \\
       Diboson & 18,\pm 1 & 18,\pm 1 \\
       Multijet & 340,\pm 170 & 470,\pm 240 \\
       \hline
       Total expected & 9860,\pm 940 & 12600,\pm 1210 \\
       \hline\noalign{\smallskip}
       Data observed & 9622, & 12706, \\
       \hline\noalign{\smallskip}
       $m_{g_{\rm Z'}}$ = 800 \gev & 200, & 224, \\
       $m_{g_{\rm KK}}$ = 1300 \gev & 59, & 65, \\
       \hline\noalign{\smallskip}
       \end{tabular}
  \end{center}
\end{table}
The reconstructed $m_{t\bar{t}}$ distribution is shown for data and background expectation
as well as two signal masses in Fig.~\ref{fig:ctrl-mass}.
\begin{table}[b!]
    \centering
\caption{Number of expected and observed events in the
 dilepton channel after applying all selection cuts
 described in Section~\ref{sec:evtsel}. The uncertainties
 shown are all normalization uncertainties as described in Section~\ref{sec:sys}.
 Statistical uncertainties on these numbers are small. }
\begin{tabular}{lD{,}{\,\,}{-1}}
\hline\noalign{\smallskip}
  & \multicolumn{1}{c}{Dilepton channel} \\
 \hline\noalign{\smallskip}
     $t\bar{t}$ 		 &  4020,\pm 470 \\ 
     Single top 		 &  210, \pm 30 \\ 
     $Z$ plus jets		 &  570, \pm 70 \\ 
     $\rm{Diboson}$ 		 &  185, \pm 30 \\ 
     $W$ plus jets and $\rm{Multijet}$ 	 	 &  190, \pm 145 \\ 
     \hline 
     Total expected 		 & 5200, \pm 500 \\ 
       \hline\noalign{\smallskip}
     Data observed		  &  5304 \\ 
     \hline 
       $m_{g_{\rm Z'}}$ = 800~\gev & 77, \\
       $m_{g_{\rm KK}}$ = 1100~\gev & 75, \\
      \hline
   \end{tabular} 
   \label{tab:dilepton_yield} 
 \end{table} 
Figure~\ref{fig:htscan} shows the \HTM~distribution for data and 
SM expectation together with a hypothetical 
KK-gluon signal with a mass of 1100~\gev\ for comparison.
(The dilepton channel has very limited sensitivity to topcolour $Z'$ bosons.) 
In both the $\ell+{\rm jets}$ and dilepton channels good 
agreement is found between data and expected background in the event yields as well as the shapes of 
kinematic distributions.

\section{Results}
\label{sec:results}

The results of this search are obtained by comparing the $m_{t\bar{t}}$ and \HTM\ 
distributions with background-only and signal-plus-background hypotheses.
The significance of a potential signal is summarized by a $p$-value, the probability of 
observing, in the absence of signal, an 
excess at least as signal-like 
as the one observed in data. The outcome of the search is ranked using the
{\sc BumpHunter}~\cite{Choudalakis:2011qn} algorithm for the $\ell+{\rm jets}$ channel and a 
likelihood ratio test statistic for the
dilepton channel. No significant deviations from SM expectations are observed.

Given the absence of a signal, upper limits are set on cross-section times branching ratio 
($\sigma \times$~BR) as a function of mass 
using a 
Bayesian approach~\cite{Bertram:2000br}.
For the limit setting, the $\ell+{\rm jets}$ channel uses variable-size binning, with bins 
ranging in size from
40 \gev\ to 500~\gev\ bins for narrow resonances, and 80 \gev\ to 500~\gev\ for 
Kaluza-Klein gluons. These values are close to the mass resolution while
limiting bin-by-bin statistical fluctuations. Mass values below 500~\gev, 
i.e.~the $t\bar{t}$ threshold region, are not considered.
A single bin contains all events with $m_{t\bar{t}} >$ 2.5 \tev.
In the dilepton channel variable-sized bins are used with bins ranging in size from 50~\gev\ 
to 200~\gev\
to maximize sensitivity while limiting bin-by-bin statistical fluctuations. The last bin contains
all events with \HTM $>$ 1.1 \tev.

The likelihood function is defined as the product of the Poisson
probabilities over all bins of the reconstructed $t\bar{t}$ invariant mass or \HTM\ 
distribution in the $\ell+{\rm jets}$ or dilepton channel, respectively. 
The Poisson probability in each bin is evaluated for the observed
number of data events given the background and signal template expectation. The total signal 
acceptance as a function of mass 
is propagated into the expectation. To calculate a likelihood for combined channels, 
the likelihoods of the individual channels 
are multiplied.

The posterior probability density is calculated using Bayes' theorem,
with a flat positive prior in the signal cross-section which is found to
be a good approximation of the reference prior \cite{Casadei:2011hx}.
Systematic uncertainties are incorporated using nuisance parameters that smear the parameters of the 
Poisson probability in each bin. For each systematic uncertainty 
a Gaussian prior controls the probability for a given deviation of the parameter from the 
nominal value. The 95\% credibility level (CL) 
upper limit on the signal cross-section times branching ratio is
identified with the 95\% point of the posterior probability. 
The expected limits are determined by using the background expectation instead of the data in the limit computation, and 
the one and two standard-deviation bands around these limits are determined from the distribution of
limits in pseudo-experiments.

Systematic uncertainties degrade the expected cross-section limits by a factor ranging
from 3.0 at low mass to 1.5 at high mass.  Of the 32 systematic uncertainties
considered, none contribute individually more than 15\% of the 
degradation.

\begin{figure}[hbtp]
\centering
\subfigure[]
{
\includegraphics[width=0.45\textwidth]{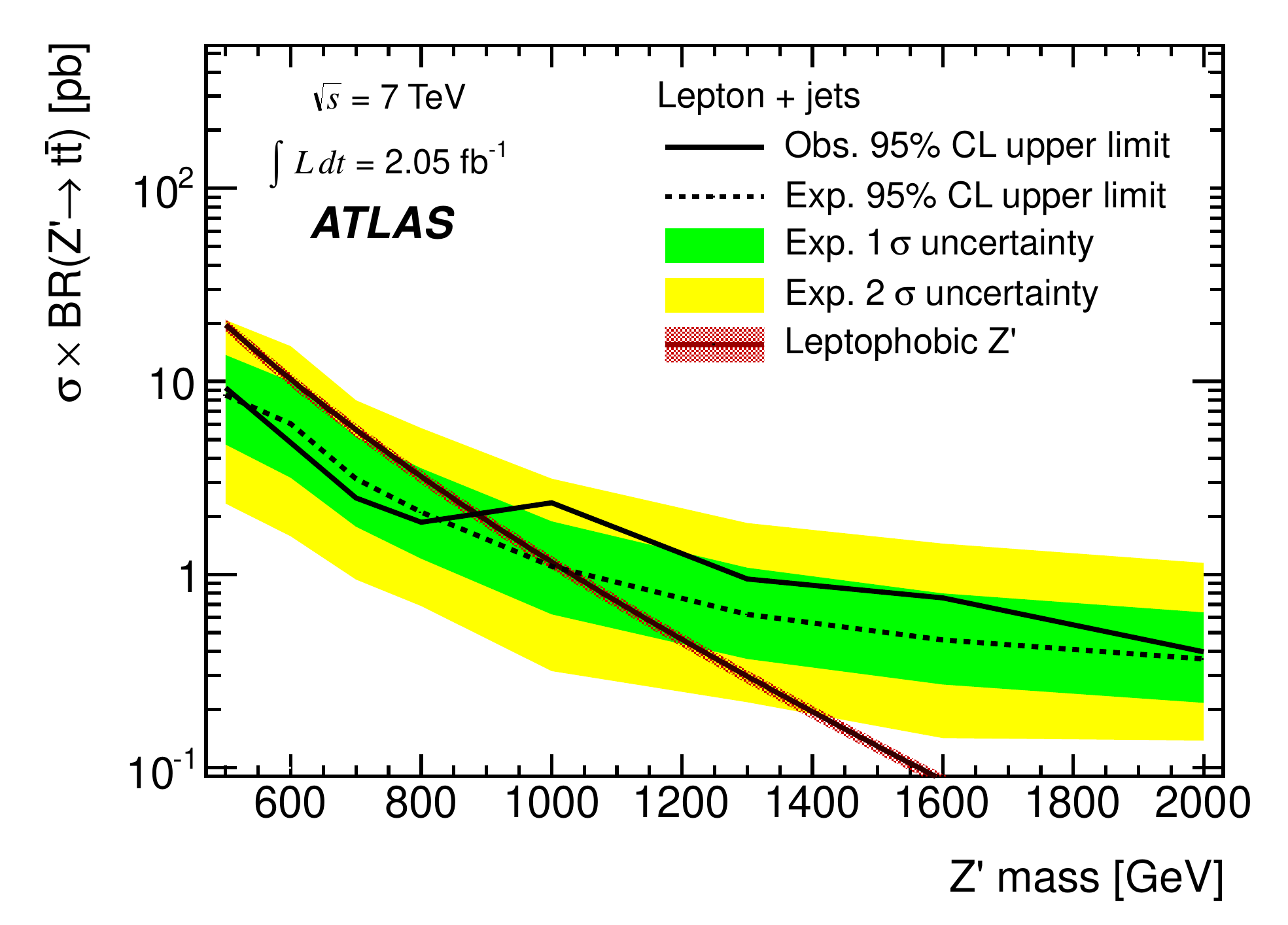}
}
\subfigure[]
{
\includegraphics[width=0.45\textwidth]{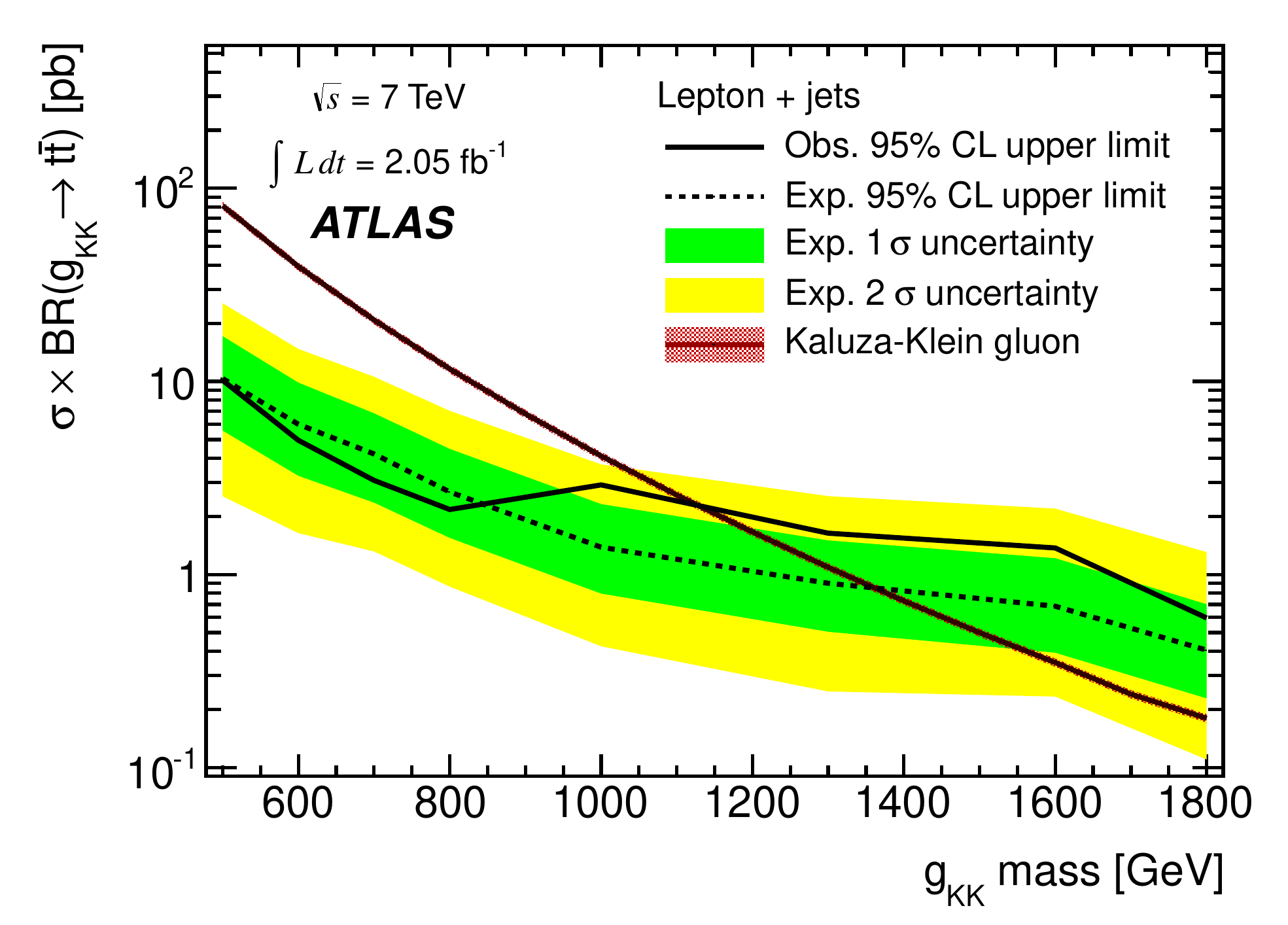}
}
\caption{Observed (solid line) and expected (dashed line) 95\% CL upper 
limits on (a) $\sigma \times$~BR($Z' \to t \bar{t}$) and (b) 
$\sigma \times$~BR($g_{\rm KK} \to t \bar{t}$) for the $\ell+{\rm jets}$ channel.
The inner and outer bands show the range in which the limit is expected to lie in 68\% and
95\% of pseudo-experiments, respectively, and the bold lines correspond to the predicted 
cross-section times branching ratio in the leptophobic topcolour and RS models.  
The bands around the signal cross-section curves
represent the effect of the PDF uncertainty on the prediction.}
\label{fig:dr-sensitivity}
\end{figure}

For the $\ell+{\rm jets}$ channel the 95\% CL observed limits on narrow and wide resonances
are shown in Fig.~\ref{fig:dr-sensitivity}, together with the predicted 
cross-section times branching ratio for the models considered and the expected 
limits.  Numerical values are given in Tables~\ref{tab:xslims} and \ref{tab:xslimsgkk}.
The observed (expected) 95\% CL limit on $\sigma \times$~BR($Z' \to t \bar{t}$) 
ranges from \limfive\ (8.5) pb at $m_{Z'} = 500$~\gev\ to \limthirteen\ (0.62) pb at  $m_{Z'} = 1300$~\gev.  The mass range
500~\gev\ $< m_{Z'} <$ \limzp\ \gev\ is excluded at 95\% CL.
The expected 
mass exclusion is 500~\gev\ $< m_{Z'} <$ 1010~\gev~\footnote{For comparison with 
the Tevatron, the observed (expected) 95\% CL exclusion limit is 
500~\gev\ $< m_{Z'} <$ \limzpold\ (930) \gev\ when using the old LO cross-section
calculation~\cite{Harris:1999ya}.}.  
The observed (expected) 95\% CL limit on $\sigma \times$~BR($g_{\rm KK} \to t \bar{t}$)
ranges from 10.1 (10.3) pb at $m_{g_{\rm KK}} = 500$~\gev\ to 1.6 (0.9) pb at  
$m_{g_{\rm KK}} = 1300$~\gev.
$g_{\rm KK}$ resonances with mass between 500 \gev\ and \limkkg\ \gev\ are excluded at 95\% CL, while
the expected mass exclusion is 500~\gev\ $< m_{g_{\rm KK}} <$ 1360~\gev.  

\begin{figure}[hbtp]
\centering
\includegraphics[width=0.45\textwidth]{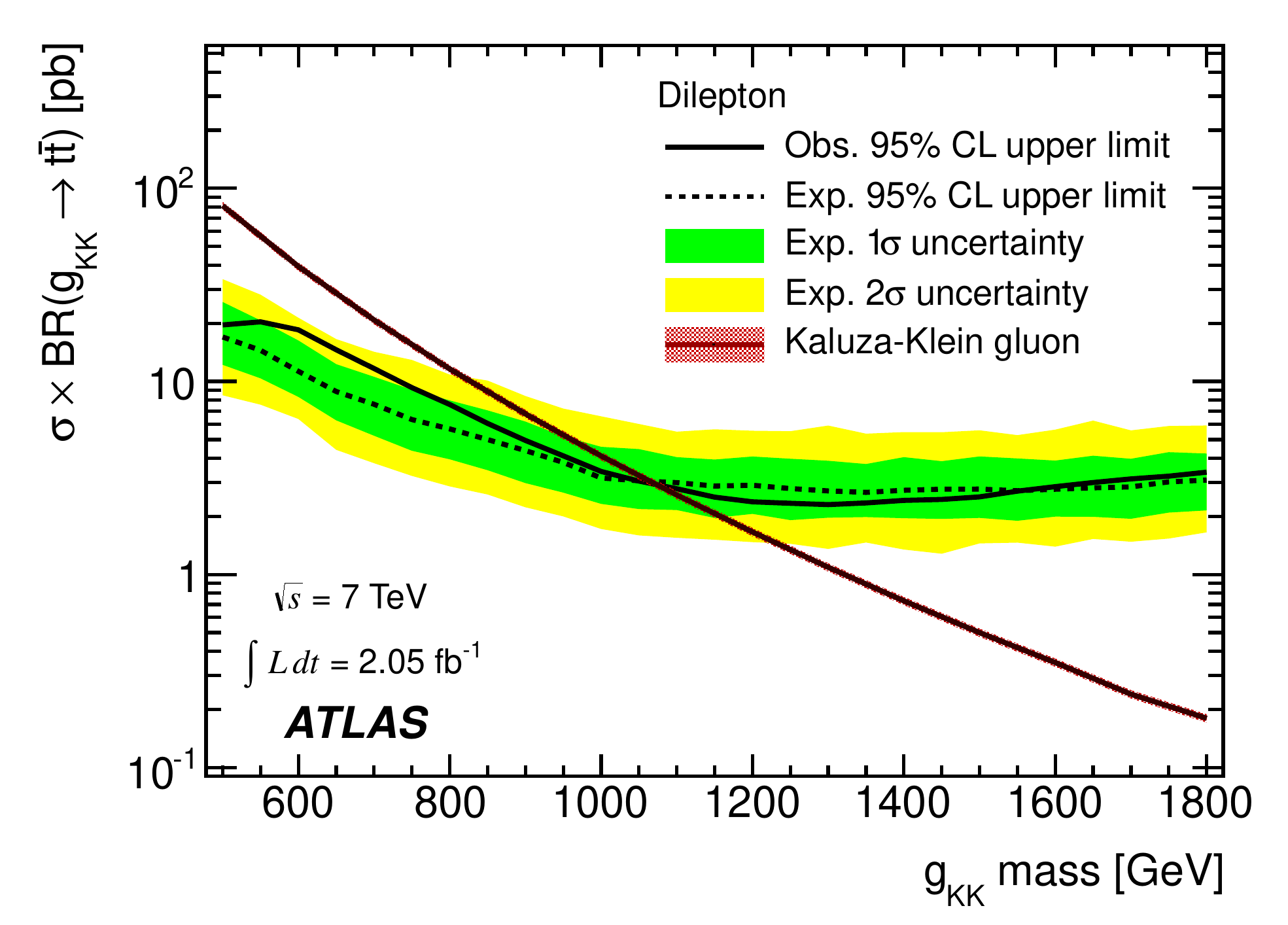}
\caption{Observed (solid line) and expected (dashed line) 95\% CL upper 
limits on $\sigma \times$~BR($g_{\rm KK} \to t \bar{t}$) for the dilepton channel.
The inner and outer bands show the range in which the limit is expected to lie in 68\% and
95\% of pseudo-experiments, respectively, and the bold line corresponds to the 
predicted cross-section times branching ratio for the RS model.
The band around the signal cross-section curve
represents the effect of the PDF uncertainty on the prediction.}
\label{fig:HTMET_limit}
\end{figure}

For the dilepton channel, the 95\% CL limits on the $g_{\rm KK}$ resonance are shown in Fig.~\ref{fig:HTMET_limit} 
with numerical values summarized in Table \ref{tab:xslimsgkk}.
The observed (expected) 95\% CL limit on $\sigma \times$~BR($g_{\rm KK} \to t \bar{t}$)
ranges from 19.6 (17.0) pb at $m_{g_{\rm KK}} = 500$~\gev\ to 2.3 (2.7) pb at  
$m_{g_{\rm KK}} = 1300$~\gev. This result excludes $g_{\rm KK}$ resonances with masses between 500 \gev\ and \limkkgDIL\ \gev\ at 95\% CL while
the expected mass exclusion is 500~\gev\ $< m_{g_{\rm KK}} <$ \limkkgDILexp\ \gev. No limit is set on $m_{Z'}$ in the dilepton channel.

Combining the $\ell+{\rm jets}$ and dilepton channels does not lead to a significant 
improvement in the limits.  However,
the dilepton channel, with different background composition and systematics, provides an
important and largely independent cross-check of the result.

\begin{table}
\begin{center}
\caption{Expected and observed 95\% CL upper limits on \mbox{$\sigma \times$~BR($Z' \to t \bar{t}$)} for
the $\ell+{\rm jets}$ channel.}
\begin{tabular}{lcc}
\hline
Mass [\gev] & $Z'$ Exp. [pb]  & $Z'$ Obs. [pb] \\ \hline
500 & 8.5 & 9.3 \\
600 & 6.0 & 4.8 \\
700 & 3.1 & 2.5 \\
800 & 2.1 & 1.9 \\
1000 & 1.1 & 2.4 \\
1300 & 0.62 & 0.95 \\
1600 & 0.46 & 0.76 \\
2000 & 0.37 & 0.40 \\ \hline
\end{tabular}
\label{tab:xslims}
\end{center}
\end{table} 
\begin{table}[h!]
\begin{center}
\caption{Expected and observed 95\% CL upper limits on $\sigma \times$~BR($g_{\rm KK} \to t \bar{t}$).}
\begin{tabular}{lcc}
\hline
\multicolumn{3}{c}{\bf $\ell+{\rm jets}$ channel} \\ \hline
Mass [\gev] & $g_{\rm KK}$ Exp. [pb]  & $g_{\rm KK}$ Obs. [pb] \\ \hline 
500 & 10.3 & 10.1 \\
600 & 6.0 & 5.0 \\
700 & 4.2 & 3.1 \\
800 & 2.7 & 2.2 \\
1000 & 1.4 & 2.9 \\
1300 & 0.90 & 1.6 \\
1600 & 0.68 & 1.4 \\
1800 & 0.41 & 0.60 \\ \hline
\multicolumn{3}{c}{\bf Dilepton channel} \\ \hline
Mass [\gev] & $g_{\rm KK}$ Exp. [pb]  & $g_{\rm KK}$ Obs. [pb] \\ \hline 
500 & 17.0 & 19.6 \\
600 & 11.3 & 18.5 \\
700 & 7.6 & 11.7 \\
800 & 5.7 & 7.6 \\
1000 & 3.2 & 3.4 \\
1300 & 2.7 & 2.3 \\
1600 & 2.8 & 2.9 \\
1800 & 3.1 & 3.4 \\ \hline
\end{tabular}
\label{tab:xslimsgkk}
\end{center}
\end{table}

\section{Summary}
\label{sec:concl}

A search for top quark pair resonances in the $\ell+{\rm jets}$ and dilepton
final states has been performed with the ATLAS experiment at the LHC.
The search uses a data sample corresponding to an integrated luminosity
of \lumitot{}, recorded at a proton-proton centre-of-mass energy of
7 \tev. The data are found to be consistent with Standard Model background expectations. 
Using the reconstructed
$t\bar{t}$ mass (\HTM) spectrum in the $\ell+{\rm jets}$ (dilepton) channel, limits are set on the 
production cross-section times branching ratio to $t\bar{t}$ for narrow and wide resonances.
In the narrow $Z'$ benchmark model, observed 95\% CL limits range from 
\limfive\ pb at $m=$ 500~\gev\ to \limthirteen\ pb at $m=$ 1300~\gev, and a 
leptophobic topcolour $Z'$ boson with 500~\gev\ $< m_{\rm Z'} <$ \limzp\ \gev\ is excluded at 95\% CL.
In the wide resonance benchmark model, Randall-Sundrum Kaluza-Klein gluons are excluded at 95\% CL 
with masses between 500 \gev\ and \limkkg\ \gev.

\section*{Acknowledgements}
\input{Acknowledgement}

\bibliographystyle{epjc}       
\bibliography{TopRes}

\clearpage
\onecolumn
\input{atlas_authlist} 
\end{document}

%% file: Acknowledgement.tex

We thank CERN for the very successful operation of the LHC, as well as the
support staff from our institutions without whom ATLAS could not be
operated efficiently.

We acknowledge the support of ANPCyT, Argentina; YerPhI, Armenia; ARC,
Australia; BMWF, Austria; ANAS, Azerbaijan; SSTC, Belarus; CNPq and FAPESP,
Brazil; NSERC, NRC and CFI, Canada; CERN; CONICYT, Chile; CAS, MOST and NSFC,
China; COLCIENCIAS, Colombia; MSMT CR, MPO CR and VSC CR, Czech Republic;
DNRF, DNSRC and Lundbeck Foundation, Denmark; EPLANET and ERC, European Union;
IN2P3-CNRS, CEA-DSM/IRFU, France; GNAS, Georgia; BMBF, DFG, HGF, MPG and AvH
Foundation, Germany; GSRT, Greece; ISF, MINERVA, GIF, DIP and Benoziyo Center,
Israel; INFN, Italy; MEXT and JSPS, Japan; CNRST, Morocco; FOM and NWO,
Netherlands; RCN, Norway; MNiSW, Poland; GRICES and FCT, Portugal; MERYS
(MECTS), Romania; MES of Russia and ROSATOM, Russian Federation; JINR; MSTD,
Serbia; MSSR, Slovakia; ARRS and MVZT, Slovenia; DST/NRF, South Africa;
MICINN, Spain; SRC and Wallenberg Foundation, Sweden; SER, SNSF and Cantons of
Bern and Geneva, Switzerland; NSC, Taiwan; TAEK, Turkey; STFC, the Royal
Society and Leverhulme Trust, United Kingdom; DOE and NSF, United States of
America.

The crucial computing support from all WLCG partners is acknowledged
gratefully, in particular from CERN and the ATLAS Tier-1 facilities at
TRIUMF (Canada), NDGF (Denmark, Norway, Sweden), CC-IN2P3 (France),
KIT/GridKA (Germany), INFN-CNAF (Italy), NL-T1 (Netherlands), PIC (Spain),
ASGC (Taiwan), RAL (UK) and BNL (USA) and in the Tier-2 facilities
worldwide.

%% file: atlas_authlist.tex
\begin{flushleft}
{\Large The ATLAS Collaboration}

\bigskip

G.~Aad$^{\rm 48}$,
B.~Abbott$^{\rm 111}$,
J.~Abdallah$^{\rm 11}$,
S.~Abdel~Khalek$^{\rm 115}$,
A.A.~Abdelalim$^{\rm 49}$,
O.~Abdinov$^{\rm 10}$,
B.~Abi$^{\rm 112}$,
M.~Abolins$^{\rm 88}$,
O.S.~AbouZeid$^{\rm 158}$,
H.~Abramowicz$^{\rm 153}$
% ...
\bigskip

% ...
$^{10}$ Institute of Physics, Azerbaijan Academy of Sciences, Baku, Azerbaijan\\
$^{11}$ Institut de F\'isica d'Altes Energies and Departament de F\'isica de la Universitat Aut\`onoma  de Barcelona and ICREA, Barcelona, Spain\\
% ...
$^{48}$ Fakult\"{a}t f\"{u}r Mathematik und Physik, Albert-Ludwigs-Universit\"{a}t, Freiburg i.Br., Germany\\
$^{49}$ Section de Physique, Universit\'e de Gen\`eve, Geneva, Switzerland\\
% ...
$^{88}$ Department of Physics and Astronomy, Michigan State University, East Lansing MI, United States of America\\
% ...
$^{111}$ Homer L. Dodge Department of Physics and Astronomy, University of Oklahoma, Norman OK, United States of America\\
$^{112}$ Department of Physics, Oklahoma State University, Stillwater OK, United States of America\\
% ...
$^{115}$ LAL, Universit\'e Paris-Sud and CNRS/IN2P3, Orsay, France\\
% ...
$^{153}$ Raymond and Beverly Sackler School of Physics and Astronomy, Tel Aviv University, Tel Aviv, Israel\\
% ...
$^{158}$ Department of Physics, University of Toronto, Toronto ON, Canada\\
% ...
\end{flushleft}


%% file: ttres.bbl
\begin{thebibliography}{10}
\providecommand{\url}[1]{\texttt{#1}}
\providecommand{\urlprefix}{URL }
\providecommand{\eprint}[2][]{\url{#2}}

\bibitem{Aad:2008zzm}
{ATLAS Collaboration}, JINST
  \href{http://dx.doi.org/10.1088/1748-0221/3/08/S08003}{\textbf{3}, S08003}
  (2008)

\bibitem{Harris:1999ya}
R.~M. Harris, C.~T. Hill, S.~J. Parke  (1999),
  \href{http://xxx.lanl.gov/abs/hep-ph/9911288}{{\tt hep-ph/9911288}}

\bibitem{Hill:1994hp}
C.~T. Hill, Phys. Lett.
  \href{http://dx.doi.org/10.1016/0370-2693(94)01660-5}{\textbf{B345}, 483}
  (1995)

\bibitem{Agashe:2006hk}
K.~Agashe, A.~Belyaev, T.~Krupovnickas, et~al., Phys. Rev.
  \href{http://dx.doi.org/10.1103/PhysRevD.77.015003}{\textbf{D77}, 015003}
  (2008)

\bibitem{Lillie:2007yh}
B.~Lillie, L.~Randall, L.-T. Wang, JHEP
  \href{http://dx.doi.org/10.1088/1126-6708/2007/09/074}{\textbf{09}, 074}
  (2007)

\bibitem{Djouadi:2007eg}
A.~Djouadi, G.~Moreau, R.~K. Singh, Nucl. Phys.
  \href{http://dx.doi.org/10.1016/j.nuclphysb.2007.12.024}{\textbf{B797}, 1}
  (2008)

\bibitem{Allanach:2009vz}
B.~C. Allanach, F.~Mahmoudi, J.~P. Skittrall, et~al., JHEP
  \href{http://dx.doi.org/10.1007/JHEP03(2010)014}{\textbf{03}, 014} (2010)

\bibitem{:2007dz}
T.~Aaltonen, et~al. (CDF), Phys. Rev. Lett.
  \href{http://dx.doi.org/10.1103/PhysRevLett.100.231801}{\textbf{100}, 231801}
  (2008)

\bibitem{:2007dia}
T.~Aaltonen, et~al. (CDF), Phys. Rev.
  \href{http://dx.doi.org/10.1103/PhysRevD.77.051102}{\textbf{D77}, 051102}
  (2008)

\bibitem{Aaltonen:2009tx}
T.~Aaltonen, et~al. (CDF), Phys. Lett.
  \href{http://dx.doi.org/10.1016/j.physletb.2010.06.036}{\textbf{B691}, 183}
  (2010)

\bibitem{Aaltonen:2011ts}
T.~Aaltonen, et~al. (CDF), Phys.Rev.
  \href{http://dx.doi.org/10.1103/PhysRevD.84.072004}{\textbf{D84}, 072004}
  (2011)

\bibitem{Aaltonen:2011vi}
T.~Aaltonen, et~al. (CDF), Phys.Rev.
  \href{http://dx.doi.org/10.1103/PhysRevD.84.072003}{\textbf{D84}, 072003}
  (2011)

\bibitem{Abazov:2008ny}
V.~Abazov, et~al. (D0), Phys. Lett.
  \href{http://dx.doi.org/10.1016/j.physletb.2008.08.027}{\textbf{B668}, 98}
  (2008)

\bibitem{Abazov:2011gv}
V.~M. Abazov, et~al. (D0), Phys.Rev.
  \href{http://dx.doi.org/10.1103/PhysRevD.85.051101}{\textbf{D85}, 051101}
  (2012)

\bibitem{cmsallhad}
{CMS Collaboration}  (2012), submitted to JHEP,
  \href{http://xxx.lanl.gov/abs/arXiv:1204.2488}{{\tt arXiv:1204.2488}}

\bibitem{Hill:1993hs}
C.~T. Hill, S.~J. Parke, Phys. Rev.
  \href{http://dx.doi.org/10.1103/PhysRevD.49.4454}{\textbf{D49}, 4454} (1994)

\bibitem{Aad:2011dr}
{ATLAS Collaboration}, Eur. Phys. J.
  \href{http://dx.doi.org/10.1140/epjc/s10052-011-1630-5}{\textbf{C71}, 1630}
  (2011)

\bibitem{LuminosityGroup:1367408}
{ATLAS Collaboration}, \emph{{Luminosity Determination in pp Collisions at
  $\sqrt{s}$ = 7 TeV using the ATLAS Detector in 2011}}, Tech. Rep.
  \href{https://cdsweb.cern.ch/record/1376384}{ATLAS-CONF-2011-116}

\bibitem{Frixione:2002ik}
S.~Frixione, B.~R. Webber, JHEP \textbf{06}, 029 (2002)

\bibitem{Frixione:2003ei}
S.~Frixione, P.~Nason, B.~R. Webber, JHEP \textbf{08}, 007 (2003)

\bibitem{Nadolsky:2008zw}
P.~M. Nadolsky, H.-L. Lai, Q.-H. Cao, et~al., Phys. Rev.
  \href{http://dx.doi.org/10.1103/PhysRevD.78.013004}{\textbf{D78}, 013004}
  (2008)

\bibitem{Corcella:2000bw}
G.~Corcella, I.~Knowles, G.~Marchesini, et~al., JHEP \textbf{0101}, 010 (2001)

\bibitem{Butterworth:1996zw}
J.~Butterworth, J.~R. Forshaw, M.~Seymour, Z. Phys.
  \href{http://dx.doi.org/10.1007/s002880050286}{\textbf{C72}, 637} (1996)

\bibitem{Moch:2008qy}
S.~Moch, P.~Uwer, Phys. Rev.
  \href{http://dx.doi.org/10.1103/PhysRevD.78.034003}{\textbf{D78}, 034003}
  (2008)

\bibitem{Kidonakis:2011wy}
N.~Kidonakis, Phys. Rev.
  \href{http://dx.doi.org/10.1103/PhysRevD.83.091503}{\textbf{D83}, 091503}
  (2011)

\bibitem{Kidonakis:2010tc}
N.~Kidonakis, Phys. Rev.
  \href{http://dx.doi.org/10.1103/PhysRevD.81.054028}{\textbf{D81}, 054028}
  (2010)

\bibitem{Kidonakis:2010ux}
N.~Kidonakis, Phys.Rev.
  \href{http://dx.doi.org/10.1103/PhysRevD.82.054018}{\textbf{D82}, 054018}
  (2010)

\bibitem{Frixione:2007vw}
S.~Frixione, P.~Nason, C.~Oleari, JHEP
  \href{http://dx.doi.org/10.1088/1126-6708/2007/11/070}{\textbf{0711}, 070}
  (2007)

\bibitem{Sjostrand:2006za}
T.~Sjostrand, S.~Mrenna, P.~Z. Skands, JHEP
  \href{http://dx.doi.org/10.1088/1126-6708/2006/05/026}{\textbf{0605}, 026}
  (2006)

\bibitem{Kersevan:2004yg}
B.~P. Kersevan, E.~Richter-Was  (2004),
  \href{http://xxx.lanl.gov/abs/hep-ph/0405247}{{\tt hep-ph/0405247}}

\bibitem{Mangano:2002ea}
M.~L. Mangano, M.~Moretti, F.~Piccinini, et~al., JHEP \textbf{0307}, 001 (2003)

\bibitem{Pumplin:2002vw}
J.~Pumplin, D.~Stump, J.~Huston, et~al., JHEP \textbf{0207}, 012 (2002)

\bibitem{Hoche:2006ph}
S.~Hoeche, F.~Krauss, N.~Lavesson, et~al.  (2006),
  \href{http://xxx.lanl.gov/abs/hep-ph/0602031}{{\tt hep-ph/0602031}}

\bibitem{Hamberg:1990np}
R.~Hamberg, W.~van Neerven, T.~Matsuura, Nucl. Phys.
  \href{http://dx.doi.org/10.1016/0550-3213(91)90064-5,
  10.1016/0550-3213(91)90064-5}{\textbf{B359}, 343} (1991)

\bibitem{Gavin:2012kw}
R.~Gavin, Y.~Li, F.~Petriello, et~al.  (2012),
  \href{http://xxx.lanl.gov/abs/arXiv:1201.5896}{{\tt arXiv:1201.5896}}

\bibitem{Sherstnev:2007nd}
A.~Sherstnev, R.~Thorne, Eur. Phys. J.
  \href{http://dx.doi.org/10.1140/epjc/s10052-008-0610-x}{\textbf{C55}, 553}
  (2008)

\bibitem{MCFM}
J.~Campbell, R.~Ellis, \emph{{MCFM---Monte Carlo for FeMtobarn processes}},
  \url{http://mcfm.fnal.gov}

\bibitem{Campbell:2011bn}
J.~M. Campbell, R.~Ellis, C.~Williams, JHEP
  \href{http://dx.doi.org/10.1007/JHEP07(2011)018}{\textbf{1107}, 018} (2011)

\bibitem{Harris:2011ez}
R.~M. Harris, S.~Jain  (2011),
  \href{http://xxx.lanl.gov/abs/arXiv:1112.4928}{{\tt arXiv:1112.4928}}

\bibitem{Gao:2010bb}
J.~Gao, C.~S. Li, B.~H. Li, et~al., Phys. Rev.
  \href{http://dx.doi.org/10.1103/PhysRevD.82.014020}{\textbf{D82}, 014020}
  (2010)

\bibitem{Alwall:2007st}
J.~Alwall, et~al., JHEP
  \href{http://dx.doi.org/10.1088/1126-6708/2007/09/028}{\textbf{0709}, 028}
  (2007)

\bibitem{Sjostrand:2007gs}
T.~Sjostrand, S.~Mrenna, P.~Z. Skands, Comput. Phys. Commun.
  \href{http://dx.doi.org/10.1016/j.cpc.2008.01.036}{\textbf{178}, 852} (2008)

\bibitem{Agostinelli:2002hh}
S.~Agostinelli, et~al. (GEANT4), Nucl. Instrum. Meth.
  \href{http://dx.doi.org/10.1016/S0168-9002(03)01368-8}{\textbf{A506}, 250}
  (2003)

\bibitem{:2010wqa}
{ATLAS Collaboration}, Eur. Phys. J.
  \href{http://dx.doi.org/10.1140/epjc/s10052-010-1429-9}{\textbf{C70}, 823}
  (2010)

\bibitem{Aad:2011mk}
{ATLAS Collaboration}, Eur. Phys. J. \textbf{C72}, 1909 (2012)

\bibitem{Aad:2012qf}
{ATLAS Collaboration}, Phys. Lett.
  \href{http://dx.doi.org/10.1016/j.physletb.2012.03.083}{\textbf{B711}, 244}
  (2012)

\bibitem{Cacciari:2005hq}
M.~Cacciari, G.~P. Salam, Phys. Lett.
  \href{http://dx.doi.org/10.1016/j.physletb.2006.08.037}{\textbf{B641}, 57}
  (2006)

\bibitem{Cacciari:2008gp}
M.~Cacciari, G.~P. Salam, G.~Soyez, JHEP
  \href{http://dx.doi.org/10.1088/1126-6708/2008/04/063}{\textbf{0804}, 063}
  (2008), \href{http://fastjet.fr/}{http://fastjet.fr/}

\bibitem{Aad:2011he}
{ATLAS Collaboration}, Submitted to Eur. Phys. J.  (2011),
  \href{http://xxx.lanl.gov/abs/arXiv:1112.6426}{{\tt arXiv:1112.6426}}

\bibitem{Collaboration:2012ef}
{ATLAS Collaboration}, Submitted to JHEP  (2012),
  \href{http://xxx.lanl.gov/abs/arXiv:1203.4606}{{\tt arXiv:1203.4606}}

\bibitem{ATLAS-CONF-2011-102}
{ATLAS Collaboration}, \emph{{Commissioning of the ATLAS high-performance
  b-tagging algorithms in the 7 TeV collision data}}, Tech. Rep.
  \href{https://cdsweb.cern.ch/record/1369219}{ATLAS-CONF-2011-102}

\bibitem{Aad:2012re}
{ATLAS Collaboration}, Eur. Phys. J.
  \href{http://dx.doi.org/10.1140/epjc/s10052-011-1844-6}{\textbf{C72}, 1844}
  (2012)

\bibitem{ATLAS:2012aa}
G.~Aad, et~al. (ATLAS Collaboration), JHEP \textbf{1205}, 059 (2012)

\bibitem{Martin:2009iq}
A.~Martin, W.~Stirling, R.~Thorne, et~al., Eur. Phys. J.
  \href{http://dx.doi.org/10.1140/epjc/s10052-009-1072-5}{\textbf{C63}, 189}
  (2009)

\bibitem{Aad:2011kp}
{ATLAS Collaboration}, Phys. Lett.
  \href{http://dx.doi.org/10.1016/j.physletb.2011.12.046}{\textbf{B707}, 418}
  (2012)

\bibitem{ATLAS-CONF-2011-089}
{ATLAS Collaboration}, \emph{Calibrating the $b$-Tag Efficiency and Mistag Rate
  in 35 pb$^{-1}$ of Data with the ATLAS Detector}, Tech. Rep.
  \href{https://cdsweb.cern.ch/record/1356198}{ATLAS-CONF-2011-089}

\bibitem{Aad:2012ug}
{ATLAS Collaboration}, Submitted to Eur. Phys. J. C.  (2012),
  \href{http://xxx.lanl.gov/abs/arXiv:1203.4211}{{\tt arXiv:1203.4211}}

\bibitem{Alwall:2007fs}
J.~Alwall, S.~Hoche, F.~Krauss, et~al., Eur. Phys. J.
  \href{http://dx.doi.org/10.1140/epjc/s10052-007-0490-5}{\textbf{C53}, 473}
  (2008)

\bibitem{Aad:2011yb}
{ATLAS Collaboration}, Phys.Lett.
  \href{http://dx.doi.org/10.1016/j.physletb.2011.12.055}{\textbf{B707}, 459}
  (2012)

\bibitem{chwalekphd}
T.~Chwalek, \emph{Ph.D. Thesis}, Tech. Rep.
  \href{http://www-ekp.physik.uni-karlsruhe.de/pub/web/thesis/iekp-ka2010-5.pdf}{IEKP-KA/2010-5}
  pp. 81-85, KIT Karlsruhe (2010)

\bibitem{Aad:2009wy}
{ATLAS Collaboration} {pp. 870--883} (2009),
  \href{http://xxx.lanl.gov/abs/arXiv:0901.0512}{{\tt arXiv:0901.0512}}

\bibitem{Choudalakis:2011qn}
G.~Choudalakis  (2011), \href{http://xxx.lanl.gov/abs/arXiv:1101.0390}{{\tt
  arXiv:1101.0390}}

\bibitem{Bertram:2000br}
I.~Bertram, et~al. (D0), \emph{{A Recipe for the construction of confidence
  limits}}, Tech. Rep.
  \href{http://lss.fnal.gov/archive/test-tm/2000/fermilab-tm-2104.shtml}{FERMILAB-TM-2104},
  Fermilab (2000)

\bibitem{Casadei:2011hx}
D.~Casadei, JINST
  \href{http://dx.doi.org/10.1088/1748-0221/7/01/P01012}{\textbf{7}, P01012}
  (2012)

\end{thebibliography}
